\documentclass[10pt,conference]{IEEEtran}
\usepackage{cite}
\usepackage{amsmath,amssymb,amsfonts}
\usepackage{mathtools, bigstrut}
\usepackage{multirow}
\usepackage{cuted}
\usepackage{lipsum}
\usepackage{algorithm}
\usepackage{algpseudocode}
\usepackage{physics}
\usepackage{caption}
\usepackage{subfig}
\usepackage{tikz}
\usepackage{algpseudocode}
\usepackage{graphicx}
\usepackage{textcomp}
\usepackage{xcolor}
\usepackage{hyperref}

	\title{ Achievable Sum-rate of  variants of QAM over  Gaussian Multiple Access Channel with and without security}
	\author{\IEEEauthorblockN{Shifa Showkat\quad Zahid Bashir Dar \quad Shahid Mehraj Shah}
 \IEEEauthorblockA{\textit{Communication Control \& Learning Lab, Dept. of Electronics and Communication Engineering} \\
National Institute of Technology Srinagar, India \\}
\{shifa\_05phd19@nitsri.net, Zahidbhat242@gmail.com, shahidshah@nitsri.ac.in\}}

\begin{document}
	\maketitle
	
	\begin{abstract}
	The performance of next generation wireless systems (5G/6G and beyond) at the physical layer is primarily driven by the choice of digital modulation techniques that are  bandwidth and power efficient, while maintaining high data rates. Achievable rates for Gaussian input and some finite constellations (BPSK/QPSK/QAM) are well studied in the literature. However, new variants of Quadrature Amplitude Modulation (QAM) such as Cross-QAM (XQAM), Star-QAM (S-QAM), Amplitude and phase shift keying (APSK), and Hexagonal Quadrature Amplitude Modulation (H-QAM) are not studied in the context of achievable rates for meeting the demand of high data rates. In this paper, we study achievable rate region for different variants of M-QAM like Cross-QAM, H-QAM, Star-QAM and APSK. We also compute mutual information corresponding to the sum rate of Gaussian Multiple Access Channel (G-MAC), for hybrid constellation scheme, e.g., user 1 transmits using Star-QAM and user 2 by H-QAM. From the results, it is observed that S-QAM gives the maximum sum-rate when users transmit same constellations. Also, it has been found that when hybrid constellation is used, the combination of Star-QAM \& H-QAM gives the maximum rate. In the next part of the paper, we consider a scenario wherein an adversary is also present at the receiver side and is trying to decode the information. We model this scenario as Gaussian Multiple Access Wiretap Channel (G-MAW-WT). We then compute the achievable secrecy sum rate of two user G-MAC-WT with discrete inputs from different variants of QAM (viz, X-QAM, H-QAM and S-QAM).It has been found that at higher values of SNR, S-QAM gives better values of SSR than the other variants. For hybrid inputs of QAM, at lower values of SNR, combination of APSK and S-QAM gives better results and at higher values of SNR, combination of HQAM and APSK gives greater value of SSR.
	\end{abstract}

\section{Introduction}
As 5G deployments begin to spread over the world, there has been a rapid rise in the number of multimedia applications, resulting in the increased demand for high data rates. Digital modulation techniques play a critical role for achieving high data rates, however, data rates, robustness to channel defects, bandwidth, and power all of them collectively influence the modulation scheme to be used for information transmission. The predicament is that all of the characteristics cannot be improved at the same time as there is a trade-off between them. Further, the optimization of these factors is purely application specific, for instance, if power efficiency is a requirement, the modulation order is to be reduced, which reduces the data rate and results in lower bandwidth efficiency. In recent years, the new variants of QAM  have been designed and analyzed for next generation wireless systems due to their high data-rates with improved bandwidth and power efficiency \cite{singya2021survey}. These new variants of QAM are square QAM , rectangular-QAM (R-QAM), cross-QAM (X-QAM), star-QAM (S-QAM), and hexagonal-QAM (H-QAM). RQAM, a case of odd power of two QAM constellations, has better rate and channel adaptation but has higher peak and average powers, as such square-QAM is usually preferred for even power of two constellations. X-QAM resolves this by removing the outer corner points of RQAM and arranging them in a cross shape leading to reduction in average energy of the constellations. The growing demand for high data rates has prompted greater research into more compact 2D constellations, leading to HQAM constellations, based on hexagonal lattice structures. HQAM, is a more power efficient two dimensional (2D) hexagonal shaped constellation for high data rates at low energy and has the densest 2D packing resulting in reduced peak and average power of the constellation.\\
 Predominantly, it is mobile users who communicate with the base station and  try to upload huge amount of data, which is also called uplink scenario in case of cellular communication and is modelled as  Multiple Access Channel (MAC) wherein multiple users communicate simultaneously with one receiver. Two user G-MAC is one of the rarest multi-user channel whose capacity region is completely characterised from information theoretic point of view \cite{liao1972coding}. For continuous and Gaussian distributed inputs, the capacity region of GMAC has been researched and is well known \cite{cover1999elements, gallager1985perspective}. However, in practical circumstances, input is usually drawn from finite alphabet set (such as M-QAM etc). In such practical cases we define constellation constrained capacity (CCC) as the maximum rate achievable by using finite constellation input. CCC of many single-user and multi-user channels has been studied rigorously \cite{harshan2011two}.It has been found that relative rotation of constellation point of one user enhances the overall sum-rate in case of multiple access channel \cite{harshan2011two,shah2020optimal}  Until now, the researchers have computed the capacity by employing the same constellations at the two inputs.  
 In the proposed work,the capacity or sum rate (SR) has been computed for the case when both users transmit mixed or hybrid variants of QAM  and the combinations giving maximum sum rate (SR) have also been computed.\\
 On the other hand physical layer security has been studied extensively recently \cite{shah2013previous}, \cite{rajesh2012secrecy}, \cite{shah2020enhancing}. Game theory based resource allocation in multi-user channels is studied in \cite{shah2016resource}, \cite{shah2017resource}.
 In\cite{mheich2014achievable}, the secrecy rate area for a broadcast channel with private messages and finite alphabet inputs has been studied. Recently, the secrecy performance of a multiple antenna wiretap channel with antenna selection assistance was explored in \cite{ouyang2019secrecy}, where the authors used binary PSK/ quadrature PSK (BPSK/QPSK) constellations as their input. Further Multiple Acccess wiretap channel has been studied extensively in the context of optimal power allocation and rate enhancement \cite{shah2012achievable}, \cite{shah2015achieving}, \cite{shah2022secrecy}.  In this paper, we have also considered a scenario wherein an adversary is  present at the receiver side and is trying to decode the information. We model this scenario as Gaussian Multiple Access Wiretap Channel (G-MAW-WT). We then computed the achievable secrecy sum rate (SSR) of two user G-MAC-WT with discrete inputs from different variants of QAM (viz, X-QAM, H-QAM and S-QAM).

\subsection{Paper Organisation}
In section II, the channel model for GMAC and its SR has been discussed.  In section III, the channel model for GMAC with eavesdropper and its SR has been discussed. The various variants of QAM and their general constellation has been discussed in section IV. In section V, SR and SSR  has been computed  for both the cases, when both the users transmit same constellations as well as when they transmit hybrid variants and comparative analysis has also been done. Finally, the paper has been concluded in section VI. 
\section{Gaussian Multiple Access Channel: No Eavesdropper case}
\begin{figure}[ht]
\centering
\includegraphics[width=7cm]{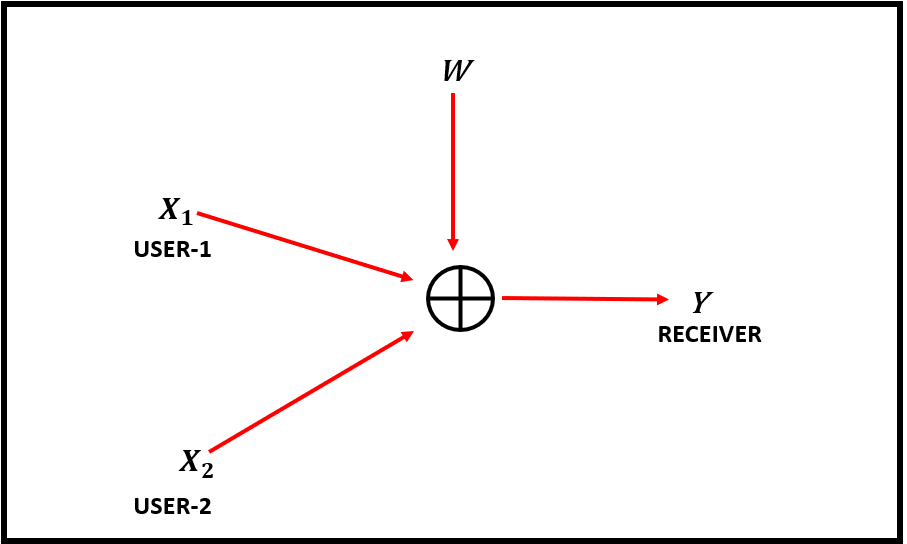}
\caption{\texttt{\small Model For GMAC Without Eavesdropper}}
\label{fig:fig1}
\end{figure}
The Gaussian Multiple Access Channel (GMAC) model is shown in Fig.1. G-MAC models a typical up-link scenario of mobile communication. The data from user-1 and user-2 are transmitted at the rates $R_1$ and $R_2$ respectively through the common communication channel and at the output, we get a noisy version of the input due to the addition of Additive White Gaussian Noise (AWGN), $W$. Both the inputs $X_1$ and $X_2$ are chosen from finite alphabet sets $\mathcal{F}_1=\{x_1,x_2,\cdots,x_{M_1}\}$ and $\mathcal{F}_2=\{x_1,x_2,\cdots,x_{M_2}\}$, having cardinalities  $M_1$ and $M_2$, respectively.
The output $Y$ at the receiver side is given as:
$Y=\left (\sqrt{P}  \right )X_1+\left ( \sqrt{P} \right )X_2 +W$ \\where $X_1 \in\mathcal{F}_1$, $X_2 \in \mathcal{F}_2$ and $W \sim \mathcal{CN}(0,\sigma^2)$ (Normally distributed with zero mean and variance $\sigma^2$).
 It has been assumed that both the inputs from finite alphabets $\mathcal{F}_1$ and $\mathcal{F}_2$ are chosen with uniform distribution, that is, all points in any set are equally probable, which implies that, $p(x_a)=\frac{1}{M_1}$ and $p(x_b)=\frac{1}{M_2}$ for $a=1,2,\cdots, M_1$ and $b=1,2,\cdots ,M_2$. It has also been assumed that the  average power constraint for  the two users is same, i.e. , $P_1=P_2=P$.\\
Let $SR= R_1 +R_2$, be the sum-rate due to both users at the receiver. The rate pairs $(R_1,R_2)$ satisfy \cite{ahlswede2014towards} 
\begin{eqnarray}
R_1 &< I(X_1;Y|X_2)\\
R_2 &< I(X_2;Y|X_1)\\
SR &< I(X_1,X_2;Y)
\label{rate_region_mac}
\end{eqnarray} 
where $I(X;Y)$ represents the mutual information. 
The upper bound of the rate pair is well known in closed form for Gaussian input.  In practice we use finite alphabet inputs hence we need to evaluate the approximate bound of these rates by computing mutual information with finite constellation. In the proposed work, we are more interested to compute the sum-rate, therefore we need to evaluate  equation for sum rate and this can be done by using chain rule of mutual information as follows:
\begin{equation}
SR=I(X_1,X_2;Y)=I(X_2;Y)+I(X_1;Y|X_2)\\
\label{Sum Rate}
\end{equation}
where $I(X_2;Y)=H(Y)-H(Y|X_2)$,\\ $H(Y)$ is the differential entropy of output and $H(Y|X_2)$ is the conditional differential entropy.
 These are defined as: 
 \begin{align}
 H(Y)=-\int_yp(y)\log_2(p(y))dy \\ H(Y|X_2)=\frac{1}{M_2}\sum^{M_2}_{r_2=1} H(Y|X_2=x_2(r_2))
 \end{align}
For the two-user GMAC, $I(X_2;Y)$ and $I(X_1;Y|X_2)$ are given in (\ref{Equation-5}) and (\ref{Equation-6}) respectively. After simplfying, we get the following form:
\begin{equation}
I(X_2;Y)=\log_2(M_2)-\frac{f_1}{M_1M_2}
\label{Equation-5}
\end{equation}

where $f_1$ is the expanded form of the following expression:
\begin{strip}
\begin{align*}
\left\{\sum\limits^{M_1}_{r_1=1}\sum\limits^{M_2}_{r_2=1}E_W\left[\log_2\left[\frac{\sum\limits^{M_1}_{t_1=1}\sum\limits^{M_2}_{t_2=1}\exp(-\left |\sqrt{P}\left (x_1(r_1)  \right )- \sqrt{P}\left (x_1(t_1)  \right )+ \sqrt{P}\left (x_2(r_2)  \right )-\sqrt{P}\left (x_2(t_2)  \right )+W  \right |^{2} /\sigma_1^2)}{\sum\limits^{M_1}_{t_1=1}\exp(-\left |\sqrt{P}\left (x_1(r_1)  \right )-\sqrt{P}\left ( x_1(t_1) \right )+W  \right |^{2}/\sigma_1^2)}\right]\right]\right\}
\end{align*}
\end{strip}

Similarly, on expanding $I(X_1;Y|X_2)$, we get the following form:
\begin{equation}
I(X_1;Y|X_2)=\log_2(M_1)-\frac{f_2}{M_1}
\label{Equation-6}
\end{equation}

where $f_2$ is the expanded form of the following expression:
\begin{tiny}
$$\sum^{M_1}_{r_1=1}E_W\left[\log_2\left[\frac{\sum\limits^{M_1}_{t_1=1}\exp(-\left | \sqrt{P}\left ( x_1(r_1) \right )- \sqrt{P}\left ( x_1(t_1) \right )+W \right |^{2}/\sigma_1^2)}{\exp\left(\frac{-|W|^2}{\sigma_1^2}\right)}\right]\right]$$
\end{tiny}
on substituting (\ref{Equation-5}) and (\ref{Equation-6}) in (\ref{Sum Rate}), we get:
\begin{equation}
SR=\log_2(M_1)+\log_2(M_2)-\frac{f_1}{M_1M_2}-\frac{f_2}{M_1}
\label{Sum Rate Expanded}
\end{equation}
 From (\ref{Sum Rate Expanded}) , it is clear that there is no closed form solution  to it, however, it can be approximated  using Monte Carlo approximation. 
 Here the expectation operators in $f_1$ and $f_2$ are with respect to the Gaussian noise W and thus, approximation is done with respect to W
 \section{Gaussian MAC with eavesdropper}
\begin{figure}[h]
\centering
\includegraphics[scale=0.6]{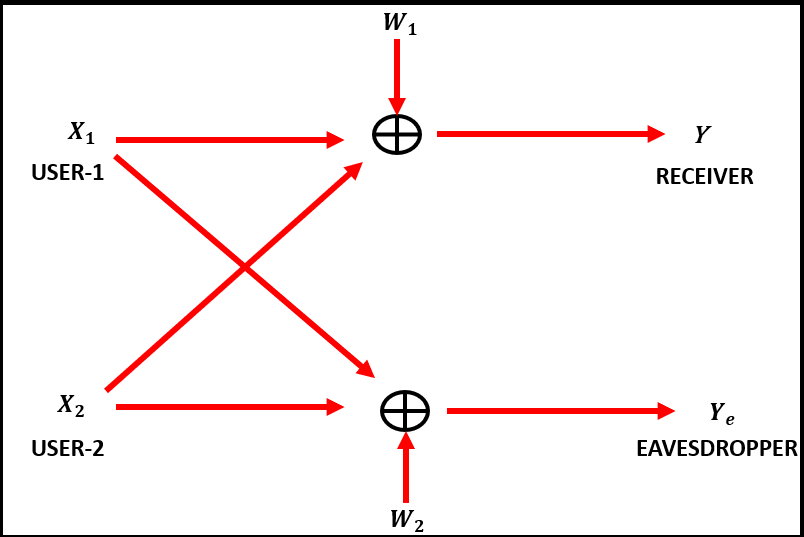}
\caption{ \texttt{\small Gaussian Multiple Access Wiretap Channel}}
\label{fig:fig2}
\end{figure}
In this channel model, there are two transmitting users with inputs, $X_{1}$ and $X_{2}$ and one legitimate receiver receiving $Y$ and and eavesdropper receiving $Y_{e}$. We assume degraded channel model (noise to Eve's channel is more than the noise of main channel). The transmitter's will employ the coding scheme for Multiple Access Wiretap Channel \cite{shah2012achievable} to secure the messages transmitted by the users from eavesdropper. As a result, the achievable rates will be reduced as compared to the model-1. 
This model can be represented as\\
\begin{eqnarray*}
Y=\left ( \sqrt{P} \right )X_1+ \left ( \sqrt{P} \right )X_2+W_1\\
Y_e=\left ( \sqrt{P} \right )X_1+\left ( \sqrt{P} \right )X_2+W_2
\end{eqnarray*}
where $W_i \sim \mathcal{CN} (0,\sigma_i^2)$ is Additive White Gaussian Noise (AWGN), for $i=1,2$ and $X_i=\{x_1,x_2,\cdots,x_M
\}$ is the channel input at transmitter i, which satisfies the power constraint i.e. $E[|X_i|^2]\leq P$.
Let $SSR= R_1 + R_2$ be the secrecy sum rate at the receiver.\\ These rates are bounded as:\\
\begin{align}
R^{'}_{1} &\leq I(X_1;Y|X_2)-I(X_1;Y_e) \nonumber\\
R^{'}_{2} &\leq I(X_2;Y|X_1)-I(X_2;Y_e) \nonumber \\
SSR &\leq I(X_1,X_2;Y)-I(X_1;Y_e)-I(X_2;Y_e)\nonumber \\
&=I(X_2;Y)+I(X_1;Y|X_2)-I(X_1;Y_e)-I(X_2;Y_e) \nonumber
\end{align}
Since we have finite constellation inputs, each term in the above equation needs to be  evaluated as :\\
Therefore,
\begin{eqnarray*}
I(X_2;Y)=H(Y)-H(Y|X_2)\\
I(X_2;Y_e)=H(Y_e)-H(Y_e|X_2)\\
I(X_1;Y_e)=H(Y_e)-H(Y_e|X_1)
\end{eqnarray*}
 We assume uniform distribution of input constellation points, i.e. probability of choosing one of the modulation points for two users is $1/M_1$ and $1/M_2$ respectively. It has also been assumed that the average power constraint for the two users is same, i.e., $P_1=P_2=P$.\\
  From the definition of entropy, it is  know that:\\
\begin{eqnarray*}
H(Y)&=&-\int p(y)\log_2(p(y))dy\\
H(Y|X_2)&=&\sum^{M_2}_{x_2=1}p(x_2)\sum_{y}p(y|x_2)log_2[p(y|x_2)]\\
H(Y|X_2)&=&\frac{1}{M_2}\sum^{M_2}_{x_2=1}H(Y|X_2=x_2)\\
\text{Similarly},\\
H(Y_e)&=&-\int p(y_e)log_2(p(y_e))dy_e\\
H(Y_e|X_2)&=&\frac{1}{M_2}\sum^{M_2}_{x_2=1}H(Y_e|X_2=x_2)\\
H(Y_e|X_1)&=&\frac{1}{M_1}\sum^{M_2}_{x_1=1}H(Y_e|X_1=x_1)\\
\end{eqnarray*}
$I(X_{1};Y_{e})$ and $I(X_{2};Y_{e}) $ are given as :\\
\begin{align*}
I(X_{1};Y_{e})&= \log _{2}\left ( M_{1} \right ) -\frac{f_{3}}{M_{1}M_{2}}\\
I(X_{2};Y_{e})&= \log _{2}\left ( M_{2} \right ) -\frac{f_{4}}{M_{1}M_{2}}
\end{align*}
where $f_{3}$ can be expanded as :

\begin{strip}
\begin{align*}
 E_{W_{2}}\left [ \log _{2} \left \{\frac{\sum_{t_{1}= 1}^{M_{1}}\sum_{t_{2}= 1}^{M_{2}} \exp\left ( -\left | \sqrt{P}\left (x _{1}\left ( r_{1} \right ) \right )-\sqrt{P}\left (x _{1}\left ( t_{1} \right ) \right )+\sqrt{P}\left (x _{2}\left ( r_{2} \right ) \right )-\sqrt{P}\left (x _{2}\left ( t_{2} \right ) \right )+ W_{2} \right |^{2}/\sigma_{2}^{2}  \right ) }{ \sum_{t_{1}=1}^{M_{2}} \exp\left ( -\left | \sqrt{P}\left (x _{2}\left ( r_{1} \right ) \right )-\sqrt{P}\left (x _{2}\left ( t_{1} \right ) \right )+W_{2} \right |^{2}/\sigma _{2}^{2} \right ) }  \right \}\right ]
\end{align*}
\end{strip}
and $f_{4}$ can be expanded as:
\begin{strip}
\begin{align*}
 E_{W_{2}}\left [ \log _{2} \left \{\frac{\sum_{t_{1}= 1}^{M_{1}}\sum_{t_{2}= 1}^{M_{2}} \exp\left ( -\left |\sqrt{P}\left (x _{1}\left ( r_{1} \right ) \right )-\sqrt{P}\left (x _{1}\left ( t_{1} \right ) \right )+\sqrt{P}\left (x _{2}\left ( r_{2} \right ) \right )-\sqrt{P}\left (x _{2}\left ( t_{2} \right ) \right )+ W_{2} \right |^{2}/\sigma_{2}^{2}  \right ) }{ \sum_{t_{1}=1}^{M_{2}} \exp\left ( -\left | \sqrt{P}\left (x _{1}\left ( r_{1} \right ) \right )-\sqrt{P}\left (x _{1}\left ( t_{1} \right ) \right )+W_{2} \right |^{2}/\sigma _{2}^{2} \right ) }  \right \}\right ]
\end{align*}
\end{strip}
Therefore on substituting, we have :
\begin{strip}
$I(X_1;Y_e)=\\
\log_2(M_1) 
-\frac{1}{M_{1}M_{2}} \sum_{r_{1}= 1}^{M_{1}}\sum_{r_{2}= 1}^{M_{2}}E_{W_{2}}\left [ \log _{2} \left \{\frac{\sum_{t_{1}= 1}^{M_{1}}\sum_{t_{2}= 1}^{M_{2}} \exp\left ( -\left |\sqrt{P}\left (x _{1}\left ( r_{1} \right ) \right )-\sqrt{P}\left (x _{1}\left ( t_{1} \right ) \right )+\sqrt{P}\left (x _{2}\left ( r_{2} \right ) \right )-\sqrt{P}\left (x _{2}\left ( t_{2} \right ) \right )+ W_{2} \right |^{2}/\sigma_{2}^{2}  \right ) }{ \sum_{t_{1}=1}^{M_{2}} \exp\left ( -\left | \sqrt{P}\left (x _{2}\left ( r_{1} \right ) \right )-\sqrt{P}\left (x _{2}\left ( t_{1} \right ) \right )+W_{2} \right |^{2}/\sigma _{2}^{2} \right ) }  \right \}\right ]$\\
$I(X_2;Y_e)=\\
\log_2(M_2) -\frac{1}{M_{1}M_{2}} \sum_{r_{1}= 1}^{M_{1}}\sum_{r_{2}= 1}^{M_{2}}E_{W_{2}}\left [ \log _{2} \left \{\frac{\sum_{t_{1}= 1}^{M_{1}}\sum_{t_{2}= 1}^{M_{2}} \exp\left ( -\left |\sqrt{P}\left (x _{1}\left ( r_{1} \right ) \right )-\sqrt{P}\left (x _{1}\left ( t_{1} \right ) \right )+\sqrt{P}\left (x _{2}\left ( r_{2} \right ) \right )-\sqrt{P}\left (x _{2}\left ( t_{2} \right ) \right )+ W_{2} \right |^{2}/\sigma_{2}^{2}  \right ) }{ \sum_{t_{1}=1}^{M_{2}} \exp\left ( -\left | \sqrt{P}\left (x _{1}\left ( r_{1} \right ) \right )-\sqrt{P}\left (x _{1}\left ( t_{1} \right ) \right )+W_{2} \right |^{2}/\sigma _{2}^{2} \right ) }  \right \}\right ]$
\end{strip}
\section{Constellations for different variants of QAM}
\subsection{Hexagonal-QAM (H-QAM)}
H-QAM has the densest 2D packing and hexagonal shaped constellation, which decreases the constellation's peak and average power, thereby making it more power efficient than other current constellations \cite{singya2021survey}. Since irregular H-QAM is very difficult to generate than regular, so we have chosen the regular H-QAM for Sum-Rate analysis, which is obtained from QAM by shifting the odd number of rows to the right by magnitude of $d$ (the side length of a hexagon). The horizontal spacing between points is $2d$ and the vertical spacing between points is $\sqrt{3}d$.\\    
The general constellation for regular H-QAM is given as\\
\vspace{3mm}
\[ 
C_{hex} =\begin{bmatrix}
-\frac{(L+1)}{2}+j\frac{(L-1)\sqrt{3}}{2} & \cdots &  \frac{(L+3)}{2}+j\frac{(L-1)\sqrt{3}}{2}\\
-\frac{(L+3)}{2}+j\frac{(L-3)\sqrt{3}}{2} &  & \frac{(L+1)}{2}+j\frac{(L-3)\sqrt{3}}{2} \\
\vdots&\cdots\cdots&\vdots\\
-\frac{(L+3)}{2}-j\frac{(L-1)\sqrt{3}}{2} & \cdots & \frac{(L+1)}{2}-j\frac{(L-1)\sqrt{3}}{2}\\
\end{bmatrix}\]
where\;\;$L=\sqrt{M}$.\\

\subsection{STAR-QAM (S-QAM)}
In this variant, both the amplitude and phase of the carrier are varied and consists of multiple concentric PSK rings with equal constellation points in each ring and identical phase angle between them. The general constellation for Star-QAM is given as\\
\[ C_{star} = \left\{ \begin{array}{ll}
         r_1e^{(jn\pi/4)}, & \mbox{ $n=0,1,\cdots,7$}\\
         \vdots&\\
         r_ce^{(jn\pi/4)}, & \mbox{$n=0,1,\cdots,7$}.\end{array} \right. \] 
where, $c$ is the number of circles/rings as there are multiple rings in S-QAM constellation diagram, $r$ represents the radius of particular ring and $n$ represents the number of points in each ring. As there are 8 points in each ring, therefore n will vary from 0 to 7.
\subsection{Cross QAM (X-QAM)}
X-QAM, also known as Cross-QAM is a modified cross-shaped constellation created by eliminating the outside corner points from square constellations and arranging them in a cross shape to lower the average energy of the constellations. It has lower peak and average power than R-QAM and offers at least 1dB improvement over R-QAM constellations.\\
For  X-QAM constellation, $M=2^{2k+1}$ with $k\geq2$
Where k is the number of bits per symbol.
 
Hence, $M= 32, 128, 512,…$ are possible constellations with X-QAM. So transmission of odd number of bits is possible. For the SR analysis, we used only 32 X-QAM at the inputs of two users. The constellation for 32 X-QAM is given as follows\\
\[ C_{32} =\begin{bmatrix}
0 & -3+j5&\cdots& 3+j5& 0 \\
-5+j3 & -3+j3 & & 3+j3 & 5+j3 \\
-5+j1& -3+j1 &\cdots&3+j1& 5+j1\\
-5-j1& -3-j1 &\cdots & 3-j1 & 5-j1 \\
-5-j3 & -1-j3 & & 3-j3& 5-j3 \\
0 & -3-j5 & \cdots& 3-j5& 0
\end{bmatrix} \]
\section{Results and Discussion}
We have computed the sum rate for different variants of QAM (viz, H-QAM, S-QAM and APSK) using Monte Carlo Simulation. 
Fig.3 shows the plot of SR for different variants of M-QAM, like 16 H-QAM, 16 APSK and 16 S-QAM. From the figure, it is clear that at lower values of SNR, 16 H-QAM gives higher rate whereas at higher values of SNR, APSK gives better results. Also in the proposed work, we have employed hybrid variants at the inputs too like a combination of S-QAM and H-QAM and computed its SR. Fig.4 shows the the plot of sum rate for different hybrid inputs of QAM i.e. 16 H-QAM and 16 S-QAM, 16 H-QAM and 16 APSK and 16 S-QAM and 16 APSK. From fig. 4, it is clear that at higher values of SNR, the combination of 16 HQAM and 16 S-QAM gives better results than the other combinations and therefore, by transmitting hybrid constellations, SR improves.
\begin{figure}[ht]
\begin{tabular}{ c  c }
 \includegraphics[width=4.5cm]{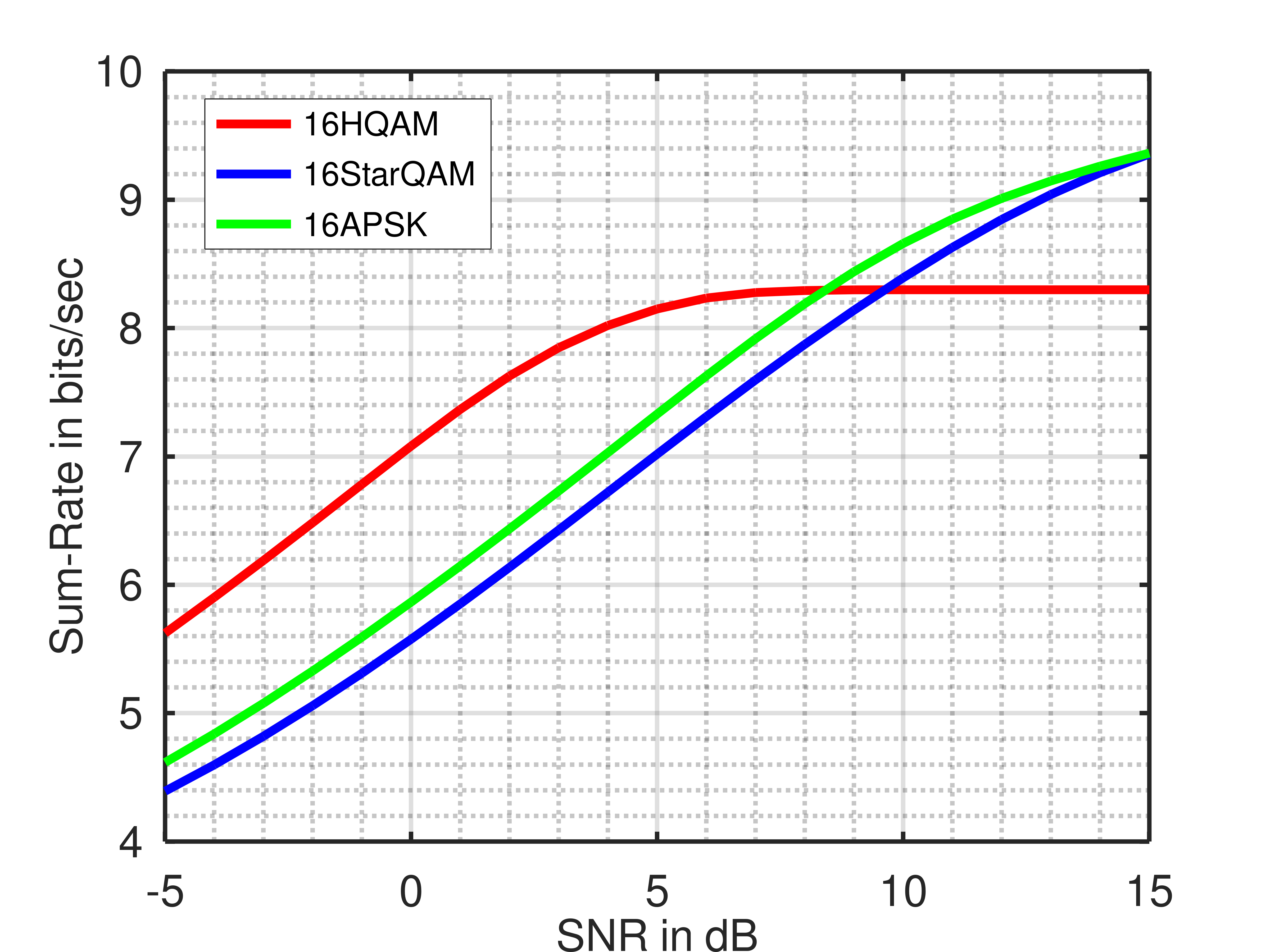}
   & \includegraphics[width=4.5cm]{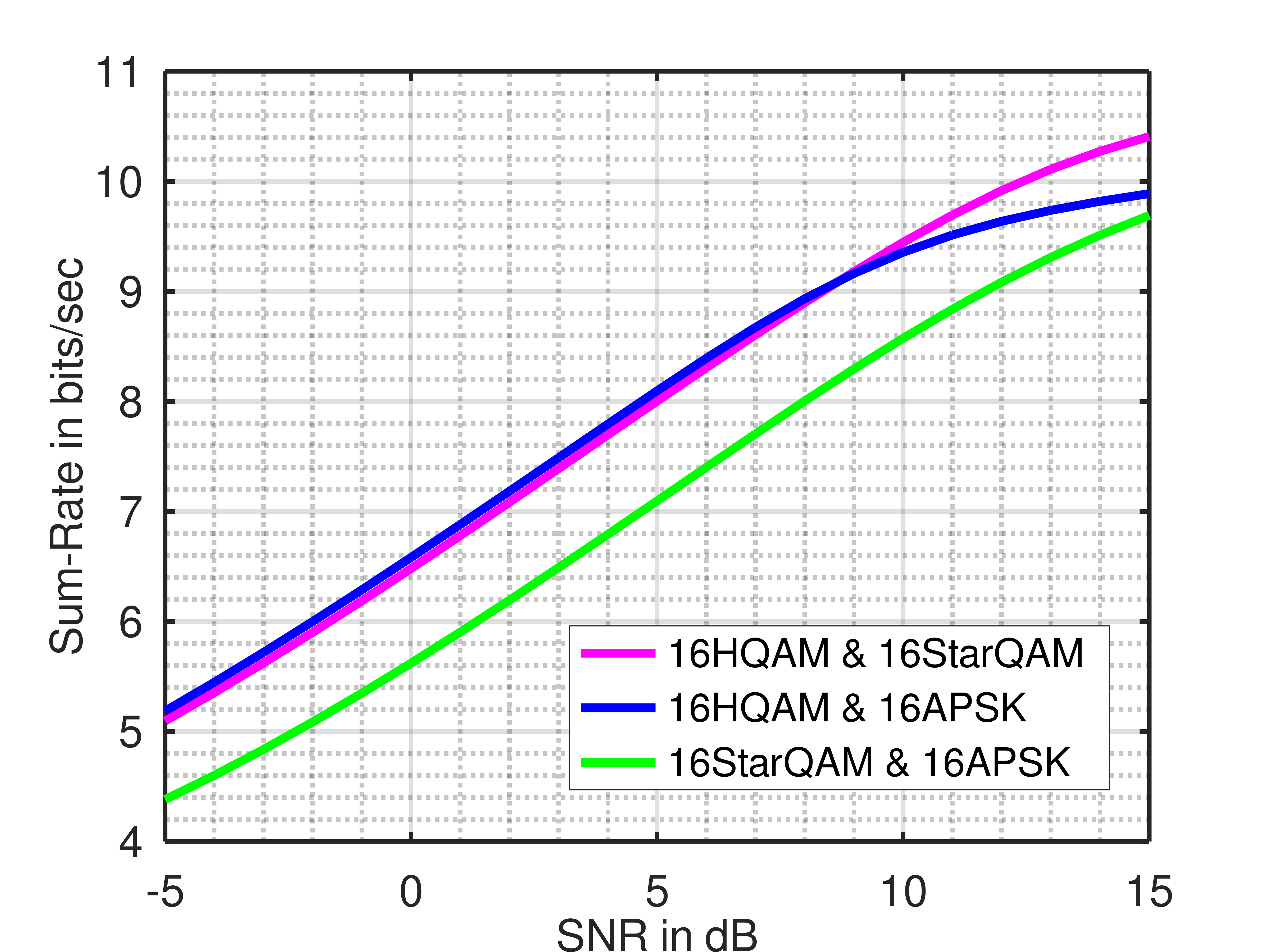}
\end{tabular}
\caption{\texttt{\small Comparison of Sum Rates For different Variants of M-QAM and Hybrid Inputs}}
\end{figure}

We have computed the Secrecy Sum Rate (SSR) for different variants of QAM like H-QAM, S-QAM and APSK by using Monte Carlo Simulation. Fig.5 shows the plot for SSR for different variants of M-QAM, from which it is clear that at higher values of SNR, S-QAM gives better values of SSR than the other variants. Also in the proposed work, we have employed hybrid variants at the inputs too like a combination of S-QAM and H-QAM and computed its secrecy sum rate. Fig.6 shows the the plot of SSR for different hybrid inputs of QAM i.e. 16 H-QAM and 16 S-QAM, 16 H-QAM and 16 APSK and 16 S-QAM and 16 APSK. From the plots, it is clear that at lower values of SNR, combination of APSK and S-QAM gives better results and at higher values of SNR, combination of H-QAM and APSK gives greater value of SSR.
\begin{figure}[ht]
    \centering
 \includegraphics[width=6cm]{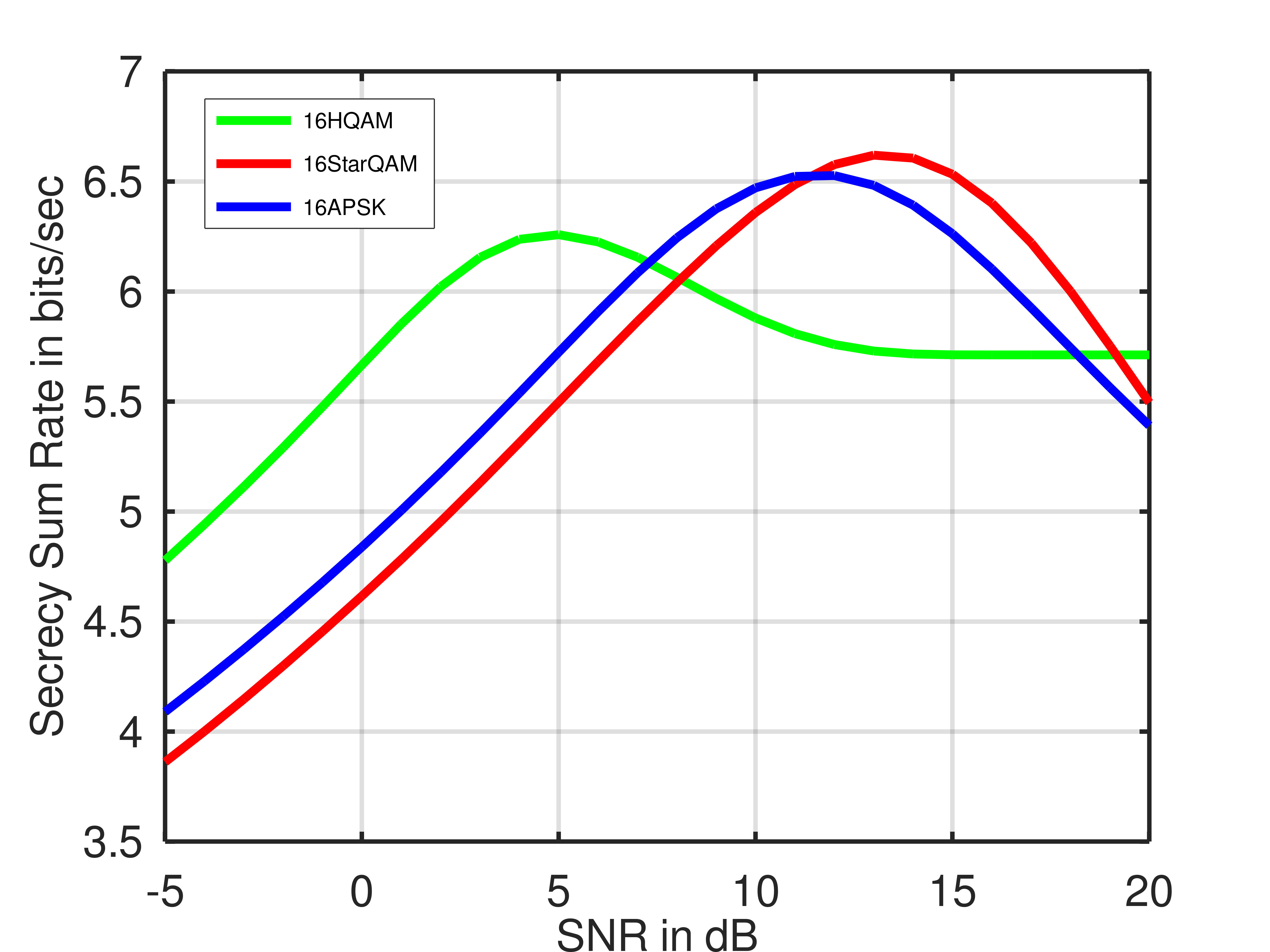}
  \caption{\texttt{\small Comparison of Secrecy Sum Rates For different Variants of M-QAM}}
    \label{fig:fig5}
    \centering
 \includegraphics[width=6cm]{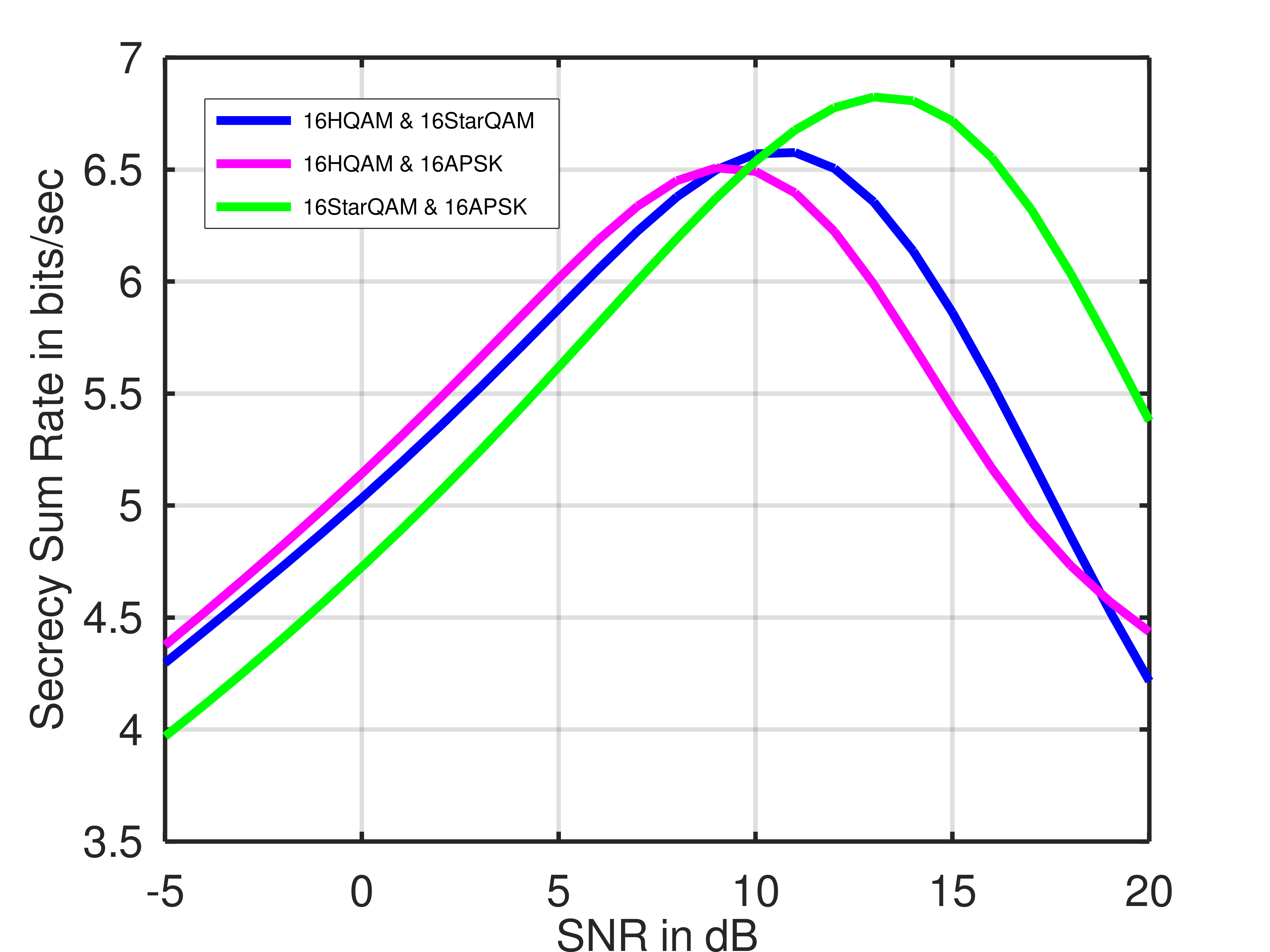}
  \caption{\texttt{\small Comparison of Secrecy  Sum Rates For Hybrid inputs of M-QAM}}
    \label{fig:fig6}
    \end{figure}
Also, it has been observed that, for some SNR range, we get an optimal angle such that if the constellation of one user is rotated by that amount, the sum-rate is enhanced. In the proposed work, we have also computed SR and SSR for different variants of M-QAM with rotation(WR) and without rotation(WOR). Fig.7, Fig.8, Fig.9 shows the sum rate for 16 H-QAM, 16 S-QAM and 32 XQAM respectively. The results have been plotted with rotation (WR) and without rotation (WOR) of constellation and from the results, it is clear that after rotation, for a particular angle, the sum rate is enhanced.
 
\begin{figure*}
\begin{minipage}[t]{0.27\textwidth}
  \includegraphics[width=\linewidth]{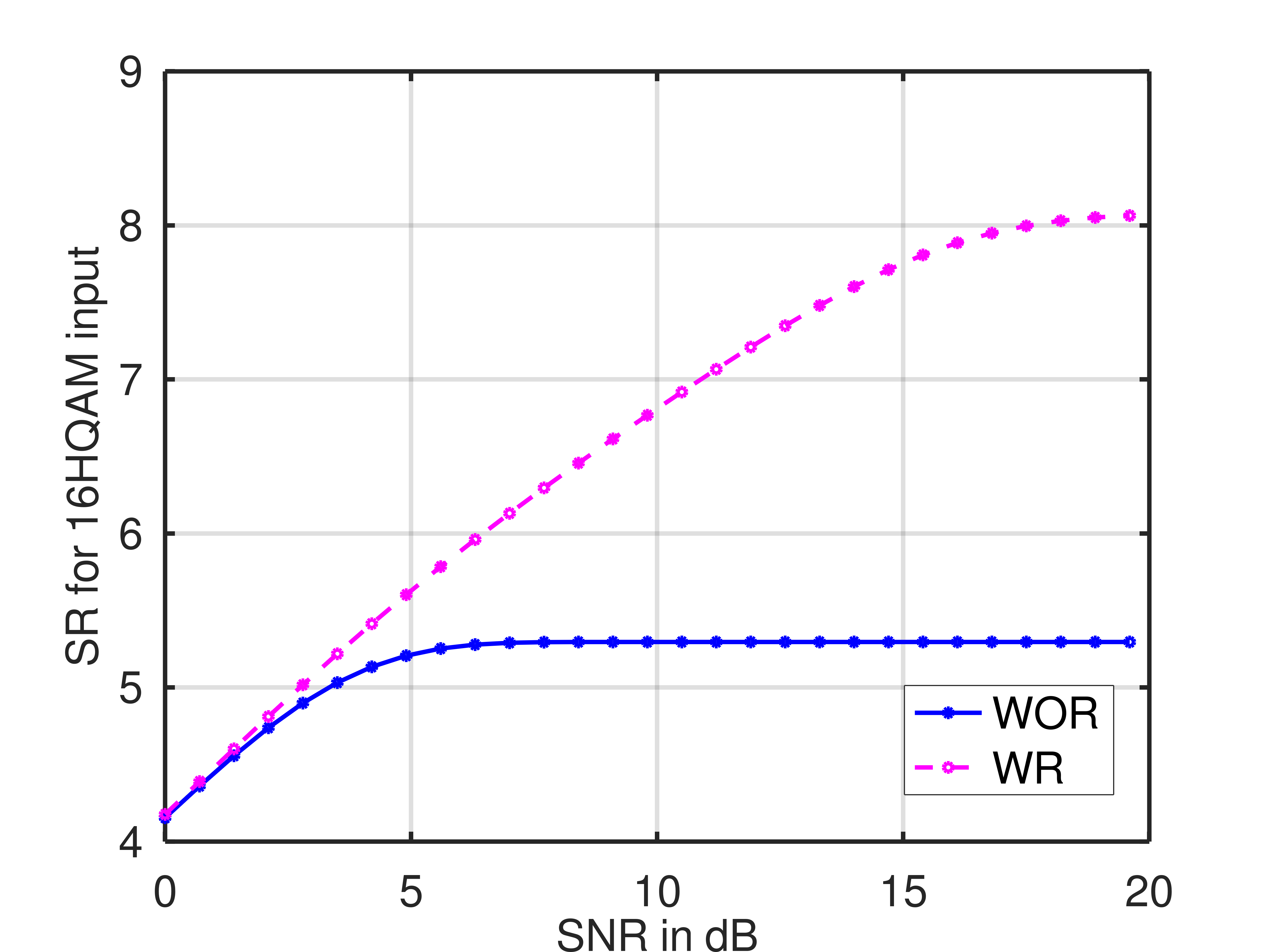}
  \caption{\texttt{\small Sum Rate for 16 HQAM input}}
  \label{fig:fig7}
\end{minipage}%
\hfill 
\begin{minipage}[t]{0.27\textwidth}
  \includegraphics[width=\linewidth]{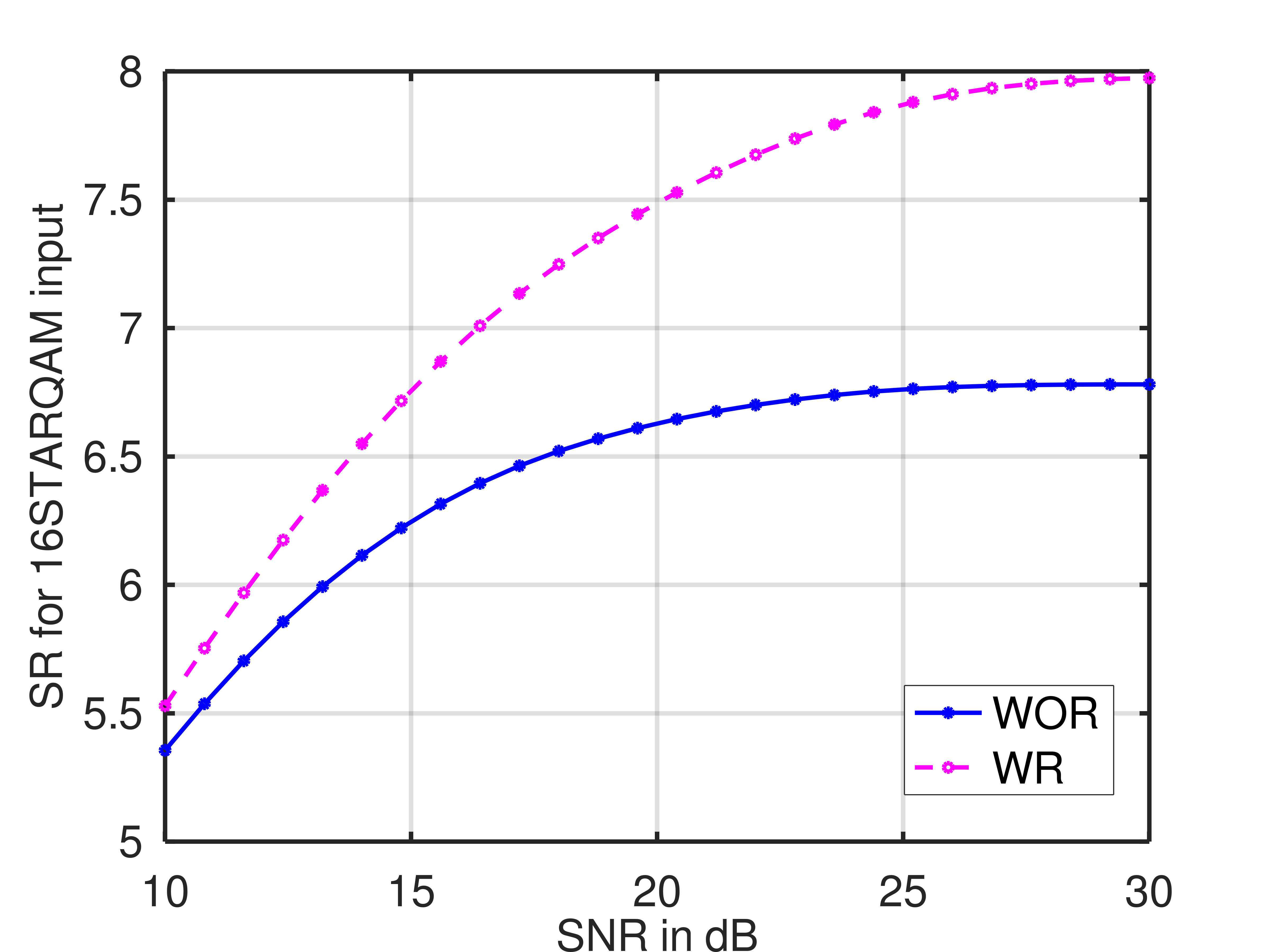}
  \caption{\texttt{\small Sum Rate for 16 STAR-QAM input}}
  \label{fig:fig8}
\end{minipage}%
\hfill
\begin{minipage}[t]{0.27\textwidth}
  \includegraphics[width=\linewidth]{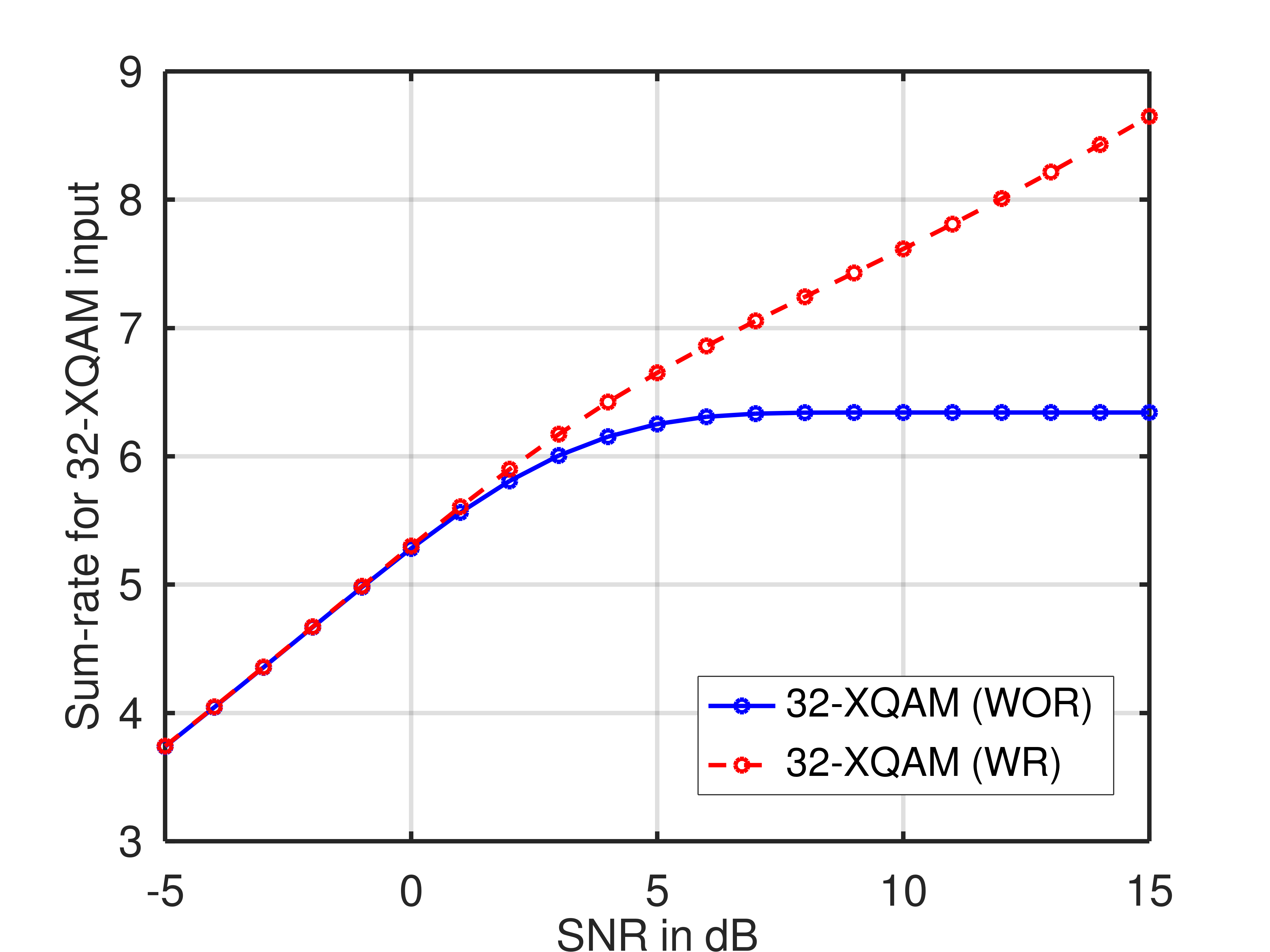}
  \caption{\texttt{\small Sum Rate for 32XQAM  input}}
\label{fig:fig9}
  \end{minipage}%
\end{figure*}
\begin{figure*}
\begin{minipage}[t]{0.27\textwidth}
  \includegraphics[width=\linewidth]{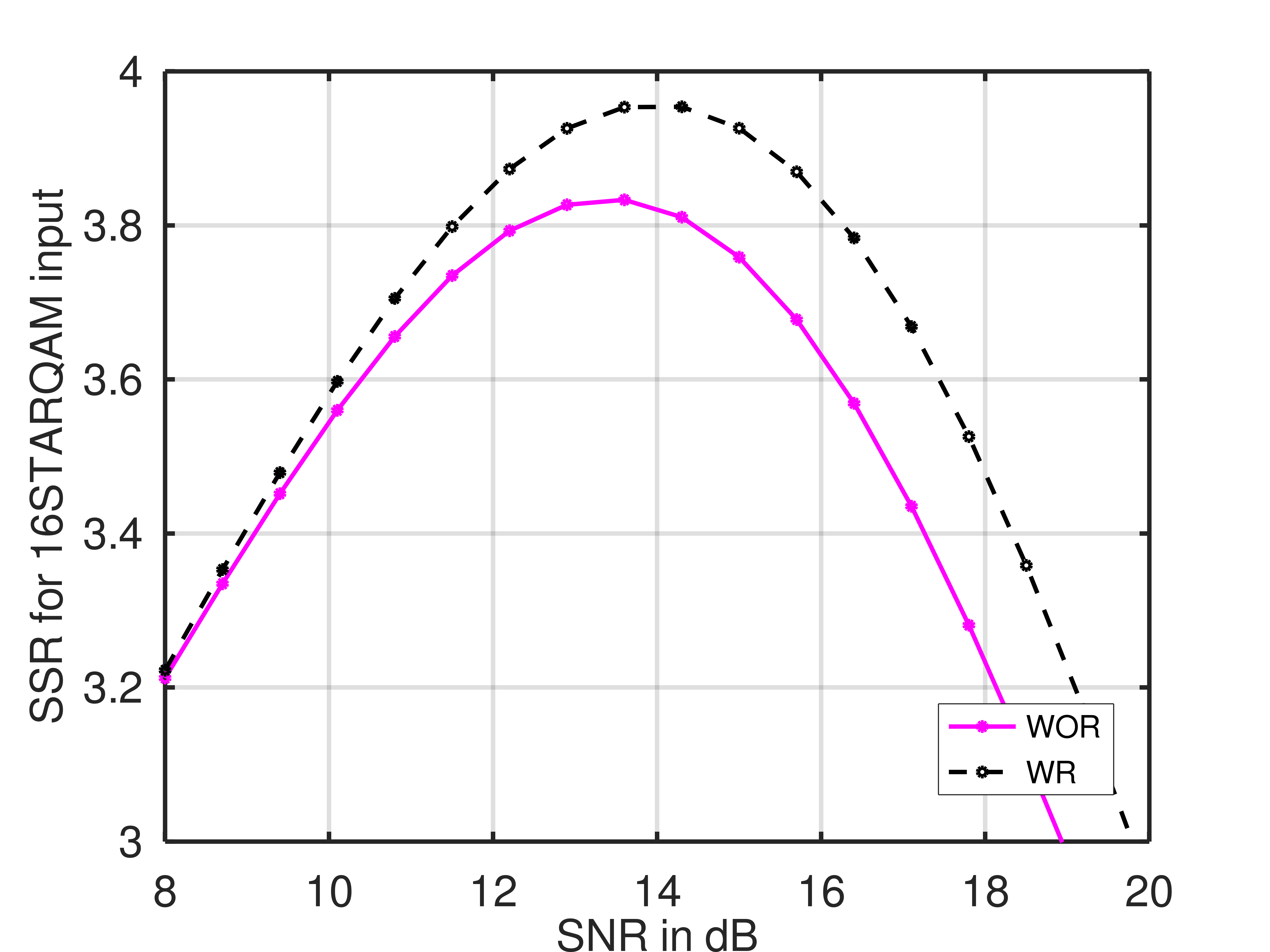}
  \caption{\texttt{\small SSR for 16STAR-QAM input}}
\label{fig:fig10}
  \end{minipage}%
\hfill 
\begin{minipage}[t]{0.27\textwidth}
  \includegraphics[width=\linewidth]{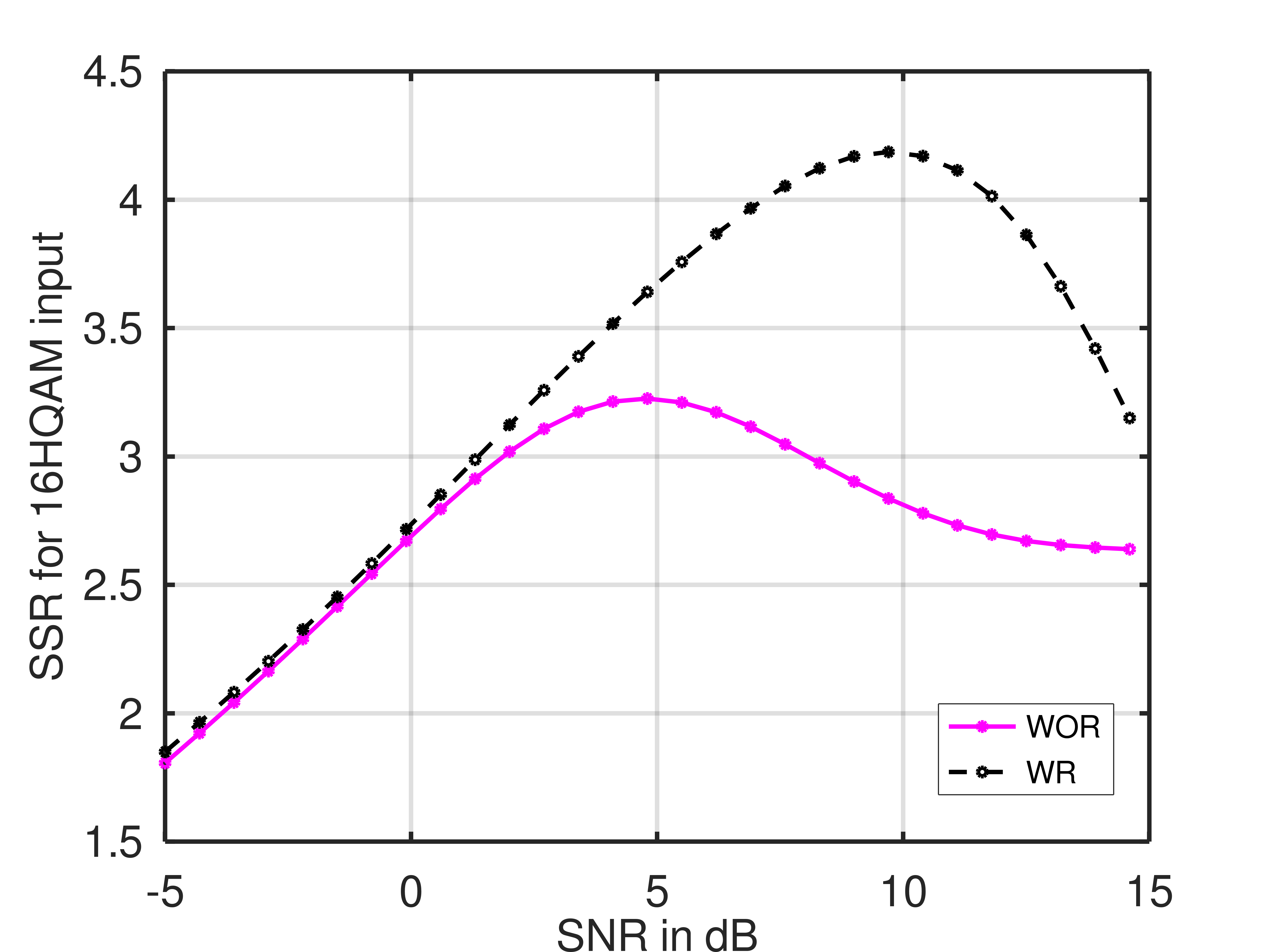}
  \caption{\texttt{\small SSR for 16HQAM input}}
\label{fig:fig11}
  \end{minipage}%
\hfill
\begin{minipage}[t]{0.27\textwidth}
  \includegraphics[width=\linewidth]{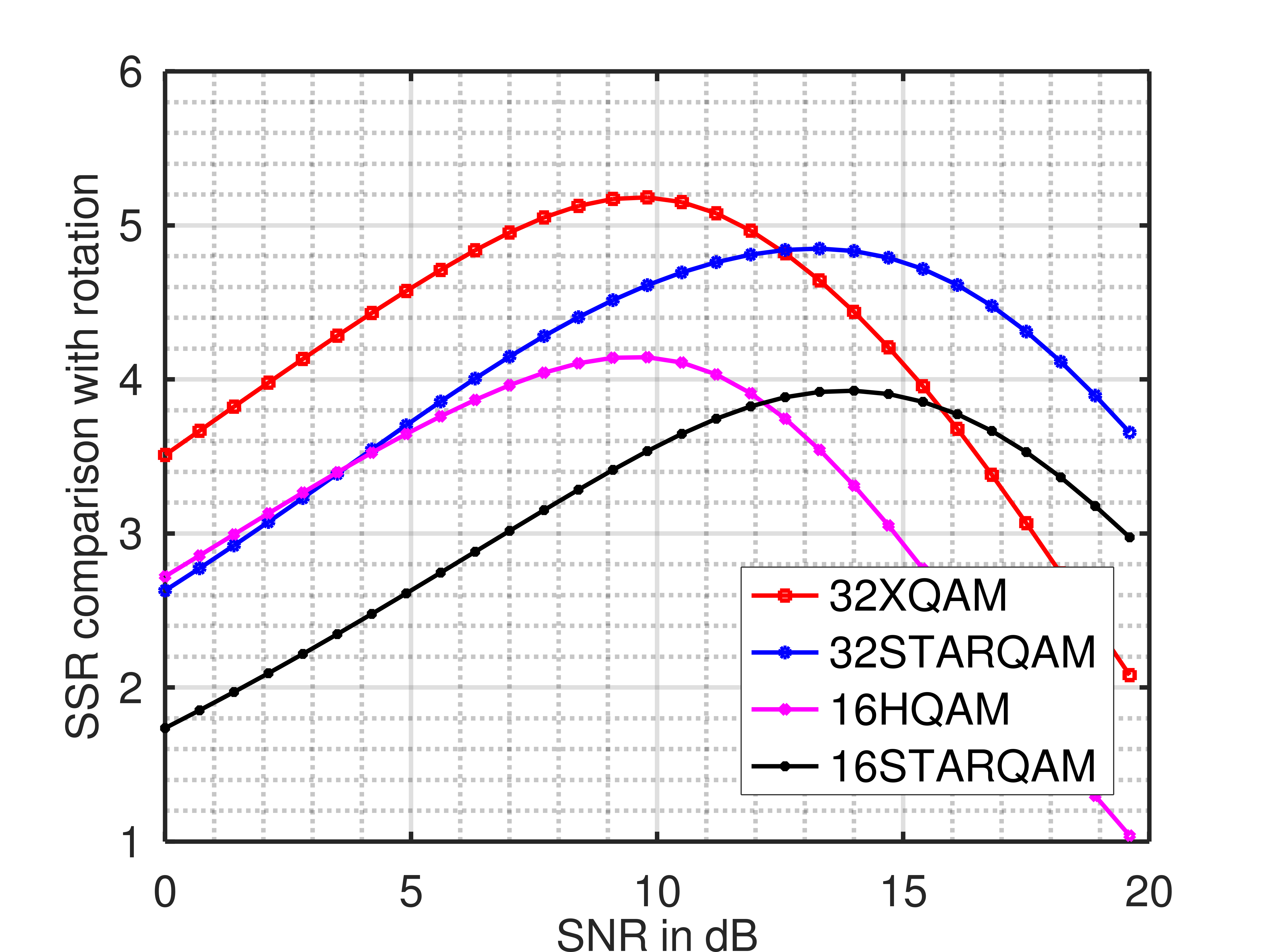}
  \caption{\texttt{\small SSR Comparison with rotation }}
\label{fig:fig12}
   \end{minipage}%
\end{figure*}

\begin{table}[htbp]
\caption{Sum Rates}
\begin{center}
\begin{tabular}{|c|c|c|c|c|}
\hline
\textbf{Modulation type} & \textbf{SNR [dB]}& \textbf{$\theta_{opt}$} & \textbf{SR(WOR)}& \textbf{SR(WR)}\\ \hline
16STARQAM & 25 & 1.2 & 6.91 & 7.95 \\ \hline
16HQAM  & 12 & 0.8 & 5.39 & 7.9\\ \hline
\end{tabular}
\label{tab1}
\end{center}
\end{table}
The SSR with rotation and without rotation  have been plotted and the results have been shown for inputs 16S-QAM and 16-HQAM in figures 10 and 11 respectively.Also a comparative SSR analysis has been done for different variants of QAM in fig.12 WR and WOR respectively.

The maximum Sum Rates and Secrecy Sum Rates obtained in both the cases are presented in Table  I and Table II.  From the results, we conclude that when users transmit same constellations, 16 S-QAM gives the maximum SR, which is further improved when they transmit hybrid constellations such as 16 S-QAM and 16 H-QAM. Also as we increase the modulation order, 32 APSK gives the maximum SR. But in case of hybrid inputs, 32 S-QAM and 32 XQAM combination gives the maximum SR.
 \begin{table}[htbp]
\caption{Secrecy Sum Rates}
\begin{center}
\resizebox{9cm}{!}{%
\begin{tabular}{|c|c|c|c|c|}
\hline
\textbf{Modulation type} & \textbf{SNR [dB]}& \textbf{$\theta^*_{opt}$} & \textbf{SSR(WOR)}& \textbf{SSR(WR)}\\ \hline
16STARQAM & 16 & 1.4 & 3.83 & 3.95 \\ \hline
16HQAM & 10 & 1.8 & 3.22 & 4.21\\ \hline
32STARQAM & 10 & 1 & 4.5 & 4.9 \\ \hline
32XQAM & 10 & 1.8 & 4.1 & 5.27\\ \hline
\end{tabular}%
}
\label{tab2}
\end{center}
\end{table}
\section{Conclusions and Future Work}
In this paper we compute the  achievable sum-rate (SR) of two-user G-MAC by employing different  variants of QAM such as X-QAM, S-QAM, APSK and H-QAM.  We also computed mutual information corresponding to the sum rate of G-MAC, while choosing different constellation for two users, e.g., user 1 transmits using S-QAM and user 2 by HQAM. The maximum $SR_{s}$ obtained in both the cases are presented in table-I and table-II. From the results, we conclude that when users transmit same constellations, 16 S-QAM gives the maximum SR, which is further improved when they transmit hybrid constellations such as 16 S-QAM and 16 H-QAM. Also as we increase the modulation order, 32 APSK gives the maximum SR and in case of hybrid inputs, combination of 32 S-QAM and 32 X-QAM gives the maximum SR. We then computed the achievable secrecy sum rate (SSR) of two user G-MAC-WT with discrete inputs from different variants of QAM (viz, X-QAM, H-QAM and S-QAM). It has been found that at higher values of SNR, S-QAM gives better values of SSR than the other variants.For hybrid inputs of QAM, at lower values of SNR, combination of APSK and S-QAM gives better results and at higher values of SNR, combination of H-QAM and APSK gives greater value of SSR. One possible direction for future research is to find approximate closed form bounds for SR and SSR for general finite constellation.

\bibliography{main.bbl}
\bibliographystyle{plain}  
\end{document}